\documentstyle{l-aa-ps}     
\input{psfig}

\begin{document}

\thesaurus{08         % A&A Section 8
              (08.02.3; % binaries: general
               08.02.6; % binaries: visual
               08.06.2; % Stars: formation
               08.16.5; % Stars: pre-main sequence
               10.15.2 Chamaeleon, Lupus, $\rho$ Ophiuchi) 
             }

   \title{Physical Properties of 90 AU to 250 AU Pre-Main-Sequence Binaries\thanks{Based on observations obtained at the European Southern Observatory, La Silla; ESO Proposal No.\ 52.7-0058, 53.7-0107, 54.D-0911 }}
%   \subtitle{}
\author{Wolfgang Brandner \inst{1} \and Hans Zinnecker \inst{2}}
 
   \offprints{Wolfgang Brandner}
 
   \institute{
Astronomisches Institut der Universit\"at W\"urzburg, Am Hubland,
D--97074 W\"urzburg, Germany\\ brandner@astro.uni-wuerzburg.de
\and 
Astrophysikalisches Institut Potsdam, An der Sternwarte 16, D--14882 Potsdam,
Germany\\
hzinnecker@aip.de}

\date{Received 11 September 1996 / Accepted 8 October 1996}
 
\maketitle

\markboth{Brandner and Zinnecker:}{Physical Properties of 90 AU to 250 AU Pre-Main-Sequence Binaries} 
 
\begin{abstract}

We have analyzed photometric and spectroscopic data of a sample of 14
spatially resolved pre-main-sequence binaries (separations
0$.\!\!^{\prime\prime}$6 to 1$.\!\!^{\prime\prime}$7)
in the nearby (150 pc) low-mass star--forming regions of Chamaeleon, Lupus, and
$\rho$ Ophiuchi. The spectroscopic data have been obtained with the
ESO New Technology Telescope (NTT) at La Silla under subarcsec seeing
conditions. All binaries (originally unresolved) were identified as 
pre-main-sequence 
stars based on their strong H$\alpha$ emission --- which classifies
them as classical T Tauri stars --- and their association with dark clouds.
One of the presumed binaries turned out to be a likely chance projection
with the ``primary'' showing neither H$\alpha$ emission nor Li absorption.

Using the spectral A index (as defined by Kirkpatrick et al.\ 1991), which
measures the strength of the CaH band at 697.5nm relative to the nearby
continuum, as a luminosity class indicator, we could show that
the classical T Tauri stars in our sample tend to be close to luminosity class 
V.

Eight out of the 14 pairs could be placed on an H--R diagram. A comparison
with theoretical pre-main-sequence evolutionary tracks yields that for
{\it all} pairs the individual components appear to be coeval within
the observational errors.
This finding is similar to Hartigan et al.\ (1994) who
detected that two third of the wider pairs with separations from 400 AU to
6000 AU are coeval.
However, unlike Hartigan et al.\ for the wider pairs, we find no non-coeval
pairs among our sample.
Thus, the formation mechanism for a significant fraction of the wider 
pre-main-sequence 
binaries might be different from that of closer pre-main-sequence
binaries. All of the latter appear to have formed simultaneously.

\end{abstract} 

\section{Introduction}

In a study entitled ``Are wide pre-main-sequence binaries coeval?\,''
Hartigan, Strom \& Strom (1994) investigated young
T~Tauri binary stars with separations $\ge$ 400 AU in order to
find clues on how binaries form. A comparison with theoretical
pre-main-sequence evolutionary tracks showed that 2/3 of the 26 binaries in 
their sample are coeval. Thus these binaries very likely formed
through fragmentation shortly before or during the collapse phase of
a molecular cloud, a process discussed, e.g., by Larson (1978), Boss (1988),
Bodenheimer et al.\ (1988), and Pringle (1989). 
In the other 1/3 of the cases, however, the secondary appears to be
considerably younger than the primary. 

Our study aimed at expanding the investigation towards closer pre-main-sequence
(PMS) binaries. Close binaries sample a different regime, because for
smaller separations circumstellar disks in binaries might
be strongly disturbed by the presence of the companion (Papaloizou
\& Pringle 1977; Artymowicz \& Lubow 1994). Thus, the PMS components of
a close binary evolve not independently from each other
due to disk mediated interaction.

We have analyzed photometric and spectroscopic data of a sample of 14
spatially resolved PMS binaries (separations
0\farcs6 to 1\farcs7, which corresponds to 90 AU to 250 AU at an average
distance of 150 pc) in the nearby low-mass star--forming regions of Chamaeleon, 
Lupus, and $\rho$ Ophiuchi. The stars had been originally identified as 
T Tauri stars based on their strong H$\alpha$ emission and their association 
with dark clouds by Schwartz (1977, desig.\ ``Sz'') and Wilking 
et al.\ (1987, desig.\ ``WSB''), and were later on resolved as 
binaries by Brandner (1992) and Reipurth \& Zinnecker (1993). 

\section{Observations and data reduction}

\begin{figure*}[ht]
\centerline{\psfig{figure=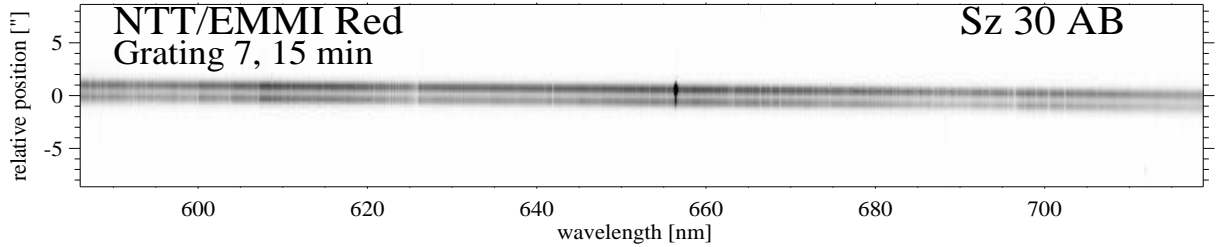,width=17.0cm}}
\caption{\label{sz30ab}
2D long-slit spectrum of the pre-main-sequence binary Sz 30
obtained with EMMI at the ESO New Technology Telescope. The strong H$\alpha$
emission lines of both components are clearly visible. 
%In spite of the
%subarcsec seeing conditions during the 15 min exposure, the wings of
%the spectra of both components are overlapping.
}
\end{figure*}

\begin{figure}[ht]
\centerline{\psfig{figure=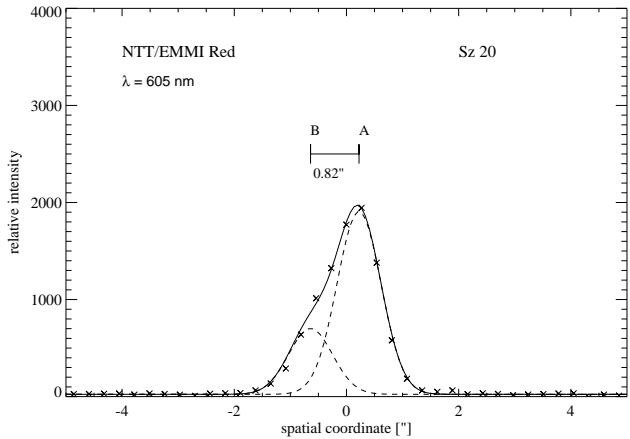,width=8.5cm,angle=90}}
\caption{\label{sz20ab}
1D cut in spatial direction of the 2D long-slit spectrum of Sz 20.
The observed intensity distribution is indicated by crosses. Two
Gaussian (dashed lines) were fitted in order to model the
intensity distribution (solid line).
}
\end{figure}

\begin{table}[ht]
\caption{\label{tthaphot}Photometric observations with the Danish 1.54m telescope/CCD camera (2.3.1995)}
\begin{center}
\begin{tabular}{ll}
object& exposure time (VRI) \\
 \hline
Sz 20      &3$\times$(10s,10s,10s)  \\
Sz 24      &3$\times$(10s,10s,10s)  \\
Sz 30      &3$\times$(10s,10s,10s)  \\
Sz 48      &3$\times$(30s,20s,20s)  \\
Sz 59      &3$\times$(10s,10s,10s)  \\
Sz 62      &3$\times$(10s,10s,10s)  \\
WSB 18     &3$\times$(30s,20s,---)  \\
WSB 19     &3$\times$(10s,10s,10s)  \\
WSB 26     &3$\times$(30s,20s,20s)  \\
\end{tabular}
\end{center}
\end{table}

\begin{table}[ht]
\caption{\label{nttspec1}Spectroscopic Observations with NTT/EMMI}
\begin{center}
\begin{tabular}{llccll}
designation& alias & $\lambda_c$ &t$_{exp}$  & sep.\ &region \\ \hline
ESO H$\alpha$281&& 660 nm & 15 min     & 1\farcs7&    Cha I \\
Sz 20      &VV Cha& 660 nm & 15 min     & 0\farcs82&  Cha I \\
Sz 24      &VW Cha& 660 nm & 10 min     & 0\farcs69&  Cha I \\
Sz 30 AB   && 660 nm & 15 min     & 1\farcs1&         Cha I \\
Sz 48      && 660 nm & 30 min     & 1\farcs5&         Cha II \\
Sz 59      &BK Cha& 660 nm & 15 min     & 0\farcs78&  Cha II \\
Sz 62      && 656 nm & 15 min     & 1\farcs14&        Cha II \\
\multicolumn{6}{c}{(20.3.1994)}\\ \hline
Sz 88      &HO Lup& 645 nm & 15 min     & 1\farcs5&   Lup 3 \\
Sz 88      &HO Lup& 415 nm & 30 min     & 1\farcs5&   Lup 3 \\
Sz 101     && 645 nm & 20 min     & 0\farcs78&        Lup 3 \\
Sz 116     && 645 nm & 10 min     & 1\farcs5&        Lup 3 \\
Sz 119     && 645 nm & 15 min     & 0\farcs59&        Lup 3 \\
WSB 18     && 645 nm & 15 min     & 1\farcs09&        $\rho$ Oph \\
WSB 19     && 645 nm & 15 min     & 1\farcs5&        $\rho$ Oph \\
WSB 26     && 645 nm & 30 min     & 1\farcs17&        $\rho$ Oph \\
\multicolumn{6}{c}{(17.5.1994)}\\
\hline
\end{tabular}
\end{center}
\end{table}

Spatially resolved VRI photometry and optical (600 nm to 720 nm)
spectroscopy were obtained at the European Southern Observatory, La Silla,
with a CCD camera attached to the Danish 1.54m telescope and with EMMI (ESO
Multi Mode Instrument) at the 3.5m New Technology Telescope.
All observations were carried out under subarcsec ($\le$0\farcs8)
seeing conditions.
The logs of the observations are given in Tables 
\ref{tthaphot} \& \ref{nttspec1}.

The data reduction was carried out using IDL, IRAF, and a stand-alone version
of GaussFit (Jeffrys et al.\ 1991). As
the spectra and images of the binary components were blended, we had to apply
special techniques in order to deblend the observed intensity distributions.
Fitting models using least-squares-fit methods proved to
produce more stable and reliable results than using highly
non-linear deconvolution techniques.

\begin{figure*}[ht]
\centerline{\psfig{figure=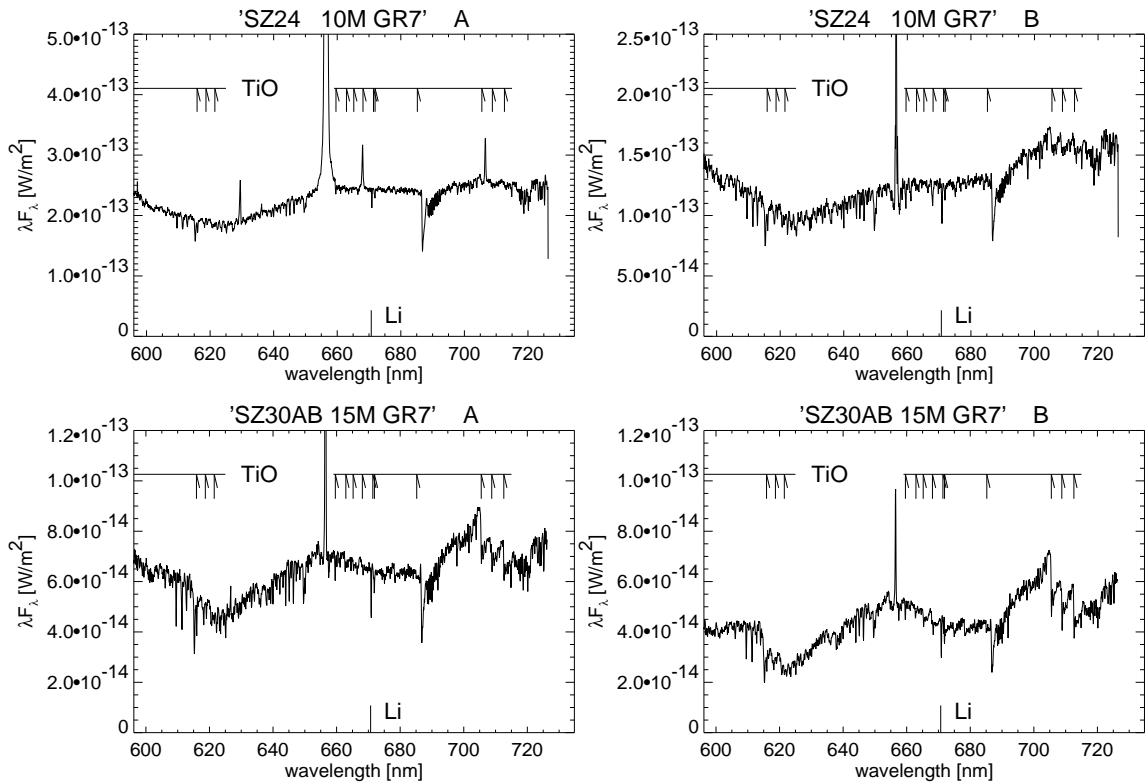,width=16.0cm,angle=90}}
\caption{\label{sz24ab}
Resolved spectra of Sz 24 (VW Cha, sep.\ 0\farcs69) and Sz 30
(sep.\ 1\farcs1). All binary components show Lithium absorption at 670.7 nm.
For Sz 24, the OI636.2nm, HeI667.9nm, and HeI706.6nm emission lines
are present only in the primary.
}
\end{figure*}

\begin{table}[ht]
\caption{\label{spectype}Comparison of derived spectral types for the
binary components with literature values for the unresolved binaries.}
\begin{center}
\begin{tabular}{lll|l}
name&\multicolumn{2}{c}{spectral type}  &  (BZ97)  \\
 \hline
VV Cha & M1 (MB80)  & M1.5 (AJK83) & M1.5 \& M3  \\
VW Cha & K5 (R80)   & K2 (AJK83) & K5/K7 \& K7 \\
Sz 30  & & M0 (H93) & M0.5 \& M2  \\
Sz 48  & & M0.5 (H$^2$92) & K7/M0 \& M0 \\
Sz 59  &M0 (H93) & K7--M0 (H$^2$92) & K5/K7 \& M0.5\\
Sz 62  &M1 (GS92) & M2 (H$^2$92) & M2 \& M3.5 \\
HO Lup & & M1 (H$^2$K$^2$94) & K7/M0 \& M2 \\
Sz 101 & & M4 (H$^2$K$^2$94) & M2.5 \& M3.5 \\
Sz 116 & & M1.5 (H$^2$K$^2$94) & M1 \& M3 \\
Sz 119 & & M4 (H$^2$K$^2$94) & M2 \& M4 \\ \hline
\end{tabular}
References: AJK83 -- Appenzeller et al.\ (1983),
BZ97 -- this paper, GS92 -- Gauvin \& Strom (1992),
H93 -- Hartigan (1993), H$^2$92 -- Hughes \& Hartigan (1992),
H$^2$K$^2$94 -- Hughes et al.\ (1994),
 MB80 -- Mundt \& Bastian (1980), R80 -- Rydgren (1980)\\
\end{center}
\end{table}

Figure \ref{sz30ab} shows as an example the 2D long-slit spectrum of the 
1\farcs1 PMS 
binary star Sz 30. The slit had been oriented along the direction of both 
components.
While the peaks of the intensity distributions for both components
are separated, the wings overlap. For even closer binaries 
the peaks are no longer resolved. This is clearly visible in Figure \ref{sz20ab},
which gives a 1D cut in spatial direction of the long-slit spectrum of the 
0\farcs82 binary Sz 20 (crosses). 
In order to separate the intensity distribution of the spectra of the binary 
components using IDL we fitted a simple model consisting of two
Gaussians (Figure \ref{sz20ab}, dashed lines) to each 1D cut in spatial direction of the 
2D spectra. The resulting resolved spectra of the 
components of Sz 24 and Sz 30 are shown in Figure \ref{sz24ab}. 
The spectra of both components of the 0\farcs69 binary Sz 24 are now clearly
separated. Note that the emission lines of OI636.2nm, HeI667.9nm, 
and HeI706.6nm are present only in the spectrum of the primary.
Sz 30 is a visual triple system with the tertiary separated by 4\farcs5
from the primary. The tertiary was not included in the present survey. 
The typical error in the 
spectra of the secondaries of the closest binaries amounts to 10\%.

For the photometric data we first modeled the local point spread function
using field stars and then applied this model to the observed
intensity distribution of the binaries using GaussFit for the 
fitting. 

The flux calibration was carried out within IRAF by comparison with
photometric and spectro-photometric standard stars\footnote{The results
of the photometric reduction and the reduced spectra of the binary
components are available from the author upon request.}.

\section{Luminosities and effective temperatures and the uncertainties involved}

\begin{figure}[ht]
\centerline{\psfig{figure=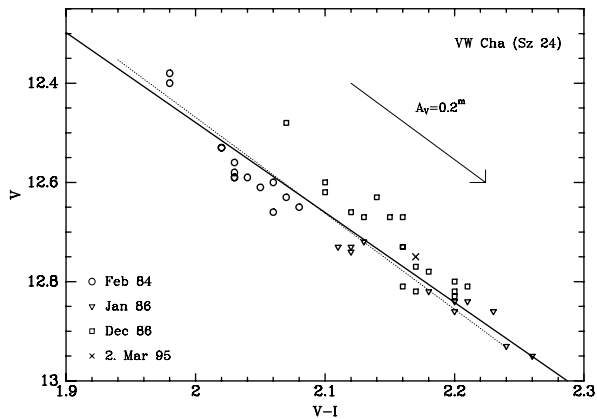,width=8.0cm,angle=90}}
\caption{\label{sz24_cmd}
Colour-magnitude-diagram for Sz 24 (VW Cha) after Bouvier et al.\
(1988). The different symbols indicate different observing runs. The
cross indicates our measurement. The solid line represents a linear
fit to the data, the dotted line is a parallel to the extinction vector.
}
\end{figure}

We used the spectral catalogues by Turnshek et al.\ (1985)
and Allen \& Strom (1995)
in order to determine the spectral types.
As spectral features like the TiO band in late-type stars are changing
rapidly with spectral type, the determination of the
spectral types should be more accurate than half a subclass. Thus, the typical
errors in the effective temperature are less than 30K.
Table \ref{spectype} compares the spectral types derived by us for the binary
components with those found in the literature for the unresolved binaries.
In general there is a good agreement. Only the early spectral classification
of VV Cha and VW Cha by Appenzeller (1977 \& 1979) was off by one
spectral class due to the veiling of weak photospheric lines in the blue
part of the spectra which had been used for spectral classification.

\begin{table}[ht]
\caption{\label{av}Comparison of values derived for the visual extinction A$_V$
of the binary components with literature values for the unresolved binaries}
\begin{center}
\begin{tabular}{l|cc|cl}
name& A& B & literature & ref.\ \\ \hline
VV Cha (Sz 20)& 0\fm43$\pm$0\fm07& 0\fm67$\pm$0\fm23 & 1\fm01 & GS92 \\
VW Cha (Sz 24)& 1\fm37$\pm$0\fm23& 1\fm26$\pm$0\fm58 & 2\fm39 &GS92 \\
Sz 30& 0\fm58$\pm$0\fm04& 0\fm19$\pm$0\fm07 & 1\fm18 &GS92 \\
Sz 48& 3\fm41$\pm$0\fm06& 3\fm58$\pm$0\fm10 & 4\fm22 &H$^2$92 \\
Sz 59& 2\fm24$\pm$0\fm08& 1\fm60$\pm$0\fm24 & 2\fm46 &H$^2$92 \\
Sz 62& 1\fm08$\pm$0\fm04& 1\fm58$\pm$0\fm11 & 1\fm54 &H$^2$92 \\ \hline
\end{tabular}
\end{center}
References: GS92 -- Gauvin \& Strom 1992, H$^2$92 -- Hughes \& Hartigan 1992
\end{table}

The spectral A index (ratio of the continuum flux at 703.5 nm to the flux 
in the CaH band at 697.5 nm) for most of our binary components is close to that 
of dwarf stars (c.f.\ Kirkpatrick et al.\ 1991). In general, due to
veiling the A index would also be closer to that of dwarf stars and hence 
could affect a better agreement than there actually is. For a moderate
veiling, however, the effect is small and can be neglected for all M-type
stars.  Thus, the surface gravity of the T Tauri stars in our sample
is close to the surface gravity of main-sequence stars and
significantly higher than the surface gravity of giants.
Walter et al.\ (1994) tried to derive the
intrinsic (R-I) and (V-K) colours of weak-line T Tauri stars by iteratively 
interpolating between the intrinsic colours of giants and of main-sequence 
stars. Due to the intrinsic IR excesses of classical T Tauri stars, this method
cannot be applied to our sample. Interestingly, Walter et al.\ (1994)
derived intrinsic (R-I) colours for T Tauri stars {\it bluer} than those
of main-sequence stars. On the other hand, synthetic colours computed
from atmospheric models of M dwarfs by Allard \& Hauschildt (1995) 
predict {\it redder} intrinsic colours for stars with a somewhat lower
surface gravity than main-sequence stars (Allard, priv.\ comm.). 
Thus, at the moment it seems
appropriate to assume colours and spectral type--effective temperature 
relations of main-sequence stars for our sample of PMS stars. We used the 
compilation provided by Hartigan et al.\ (1994).
They thoroughly discussed various sources of errors 
and uncertainties in estimating luminosities and effective temperatures for PMS 
binary components and concluded that variability of the stars is the main
source of error. 

For VW Cha (Sz 24), the brightest star in our sample, more than
300 photometric measurements are available in the literature, half of them 
in the Bessell/Cousins broad-band photometric system. These measurements
yield a variability of VW Cha from V=12\fm3 to 13\fm0.
A colour-magnitude-diagram based on 43 
measurements by Bouvier et al.\ (1988) indicates that the 
variability is conform with variable extinction (Figure \ref{sz24_cmd}). 
Variable extinction has been proposed by Grinin (1992) as one
possible source for the variability observed in many young stars.
It might explain the majority of the photometric variability between 500 nm 
and 1 $\mu$m. In this wavelength region, the largest contribution to the
overall spectral energy distribution (SED) usually is from the
the stellar photosphere (Bertout et al.\ 1988;
Kenyon \& Hartman 1990). On the other hand, among extreme CTTS veiling 
dominates the SED even at these wavelenghts. For extreme CTTS a similar
trend of V vs.\ V-I may be caused by variations in the veiling (a stronger
veiling makes the star appear bluer and brighter). 
In the blue (U, B band) for almost all T Tauri stars
the hot boundary layer, hot spots on the stellar surface, and/or photospheric
light scattered by circumstellar material contribute more to the overall SED.
Hence, the observed variability in U and B cannot be explained by
variable extinction alone.

The scatter of 0\fm05 in brightness of the unresolved binary VW Cha around 
the linear fit in Figure \ref{sz24_cmd} corresponds to a scatter of 0\fm035 per component, 
which is considerably less than the scatter of 0\fm17 assumed by Hartigan 
et al.\ (1994) to be typical for T Tauri stars in the H band.

Assuming that variable extinction is the main course of variability for
all stars in our sample the uncertainty in the relative
photometry of the components in each binary is the largest source of error
in estimating the stellar luminosity. It amounts to up to 0\fm2 in V
for the companion of Sz 24, the binary with the closest separation (0\farcs69) 
for which we were able to obtain spatially resolved photometry.
In general, however, the errors are considerably
smaller (0\fm05). We note that this has not to be true for some
extreme T Tauri stars like AA Tau, where variations of 1 mag are believed
to occur due to hot star spots (Hartigan et al.\ 1991).
The uncertainties resulting from the relative photometry
of the binary components are correlated: a fainter secondary means a brighter
primary and vice versa. 

\begin{table*}[ht]
\caption{\label{tthahrd} Physical properties of the close binary components.
Masses and ages are derived from PMS evolutionary tracks by D'Antona \& 
Mazzitelli (1994) based on opacities by Alexander et al.\ (1989) and
the Canuto \& Mazzitelli (1990) description of convection.}
\begin{center}
\begin{tabular}{l|cc|cc|cc|cc|cc|cc|}
&\multicolumn{2}{c|}{ESO H$\alpha$ 281} & \multicolumn{2}{c|}{ Sz 20} &
\multicolumn{2}{c|}{Sz 24} & \multicolumn{2}{c|}{Sz 30} &
\multicolumn{2}{c|}{Sz 48} & \multicolumn{2}{c|}{Sz 59} \\
 & A & B & A & B & A & B & A & B & A & B & A & B \\ \hline
separation
&\multicolumn{2}{c|}{1\farcs7} & \multicolumn{2}{c|}{0\farcs82} &
\multicolumn{2}{c|}{0\farcs69} & \multicolumn{2}{c|}{1\farcs10} &
\multicolumn{2}{c|}{1\farcs5} & \multicolumn{2}{c|}{0\farcs78} \\
SpT
& M0.5 & M4.5 & M1.5& M3 & K5/K7 & K7 & M0.5 & M2 & K7/M0 & M0 & K5/K7 & M0.5 \\
T$_{\rm {eff}}$ [K]
& 3730 & 3085 & 3570& 3335& 4185 & 4045 & 3730 & 3490 & 3925 & 3820 & 4185 & 3730 \\
A index
& 1.04 & 1.24 & 1.15 & 1.26 & 1.02 & 1.09 & 1.10 & 1.15 & 1.10 & 1.09 & 
%Sz59A
  1.02 & 1.10 \\ 
A$_V$ [mag] &&      & 0.43 & 0.67 & 1.37 & 1.23 & 0.58 & 0.19  & 3.41 & 3.58 &
  2.24 & 1.60 \\
L$_{\mbox{bol}}$ [L$_\odot$] &
      &      & 0.092 & 0.076 &0.61  & 0.49 & 0.23 & 0.16  & 0.096 & 0.063 &
0.16  &0.074\\
age [yr] & & & 6$\cdot$10$^6$ & 3$\cdot$10$^6$ &2$\cdot$10$^6$  & 2$\cdot$10$^6$  & 2$\cdot$10$^6$ & 2$\cdot$10$^6$  & 1$\cdot$10$^7$ & 1.5$\cdot$10$^7$ &
1$\cdot$10$^7$  &9$\cdot$10$^6$ \\
M [M$_\odot$]
 &   & & 0.35 & 0.20 &0.70  & 0.60 & 0.48 & 0.30  & 0.65 & 0.56 &
0.78  &0.40 \\ 
M$_B$/M$_A$
&\multicolumn{2}{c|}{} & \multicolumn{2}{c|}{0.57} &
\multicolumn{2}{c|}{0.86} & \multicolumn{2}{c|}{0.63} &
\multicolumn{2}{c|}{0.86} & \multicolumn{2}{c|}{0.51} \\
E$_{H\alpha}$ [nm]
 & 0.11  &-2.3 & -9.4 & -1.1 &-7.9  & -0.48 & -1.1 & -0.26  & -1.4 & -3.9 &
-5.4  &-2.9 \\ 
E$_{LiI}$ [pm]
 & ---  &27 & 43 & 74 &26  & 59 & 62 & 75  & 75 & 64 &
43  &72 \\ 
lg N$_{Li\,I}$
& --- & ? & -9.0? & -8.8? & -9.4 & -8.6 & -8.4 & -8.5? & -7.9 & -8.3 & 
%Sz59A
  -9.0 & -8.2\\ 
\hline \hline
&
\multicolumn{2}{c|}{Sz 62}  & \multicolumn{2}{c|}{Sz 88} &
\multicolumn{2}{c|}{Sz 101} & \multicolumn{2}{c|}{Sz 116} &
\multicolumn{2}{c|}{Sz 119} & \multicolumn{2}{c|}{WSB 18} \\
& A & B & A & B & A & B & A & B & A & B & A & B  \\ \hline
separation
&
\multicolumn{2}{c|}{1\farcs14} & \multicolumn{2}{c|}{1\farcs5} &
\multicolumn{2}{c|}{0\farcs78} & \multicolumn{2}{c|}{1\farcs5} &
\multicolumn{2}{c|}{0\farcs59} & \multicolumn{2}{c|}{1\farcs09} \\
SpT
%Sz88A
&M2 & M3.5 & K7/M0 & M2 & M2.5 & M3.5 & M1 & M3 & M2 & M4 &
%WSB18
  M2 & M2.5  \\
T$_{\rm {eff}}$ [K]
%Sz88A
&3490 & 3255 & 3925 & 3490& 3415& 3255& 3645& 3335& 3490& 3170 &
%WSB18
 3490 &3415\\
A index
& 1.17 & 1.30 & 
%Sz88A
  1.08 & 1.28 & 1.30 & 1.19 & 1.17 & 1.32 & 
%Sz119
  1.22 & 1.32 & 1.30 & 1.22  \\
A$_V$ [mag] &1.08 & 1.58 &&      &      &      &      &      & 
       &      & 4.04 &3.41   \\
L$_{\mbox{bol}}$ [L$_\odot$] &
0.12  &  0.074 &      &     &       &       &      &      &      &       &0.16 & 0.062 \\
age [a]&1.5$\cdot$10$^6$ & 1.5$\cdot$10$^6$ & &   &   &  &  &  &  &   & 2$\cdot$10$^6$ & 5$\cdot$10$^6$ \\
M [M$_\odot$] &0.29 & 0.16 & &   &   &  &  &  &  &   & 0.30 & 0.24 \\
M$_B$/M$_A$&
\multicolumn{2}{c|}{0.55} &\multicolumn{2}{c|}{} & \multicolumn{2}{c|}{} &
\multicolumn{2}{c|}{} & \multicolumn{2}{c|}{} &
\multicolumn{2}{c|}{0.80} \\
E$_{H\alpha}$ [nm]
 &-26.4 & -7.0 & $<$-13  &-3.4 & -0.93 & -4.4 &-0.41  & -0.48 & -0.41 & -0.46  & -0.84 & -14.0 \\
E$_{LiI}$ [pm]
 & 50 & 68& 29 &68 & 85 & 10 &45  & 29 & 69 & 170  & 75 & 28 \\
lg N$_{Li\,I}$
& -9.0? & -8.8? & 
%Sz88A
  -9.7 & -8.5-8.4? & -8.4? & ? & -9.0? & ? & 
%Sz119
  -8.6? & -7.9?? & ? & ? \\
\hline \hline
& \multicolumn{2}{c|}{WSB 19} & \multicolumn{2}{c|}{WSB 26} \\
& A & B & A & B \\ \hline
separation & \multicolumn{2}{c|}{1\farcs5} & \multicolumn{2}{c|}{1\farcs17} \\
SpT & M3 & M3.5 & M0 & M3 \\
T$_{\rm {eff}}$ [K] & 3335&3255& 3820 &3335\\
A index & 1.24 & 1.30 & 1.14 & 1.23 \\
A$_V$ [mag] & 1.81 & 2.18 & &  \\
L$_{\mbox{bol}}$ [L$_\odot$] & 0.095 &0.061&       &         \\
age [a]& 2.5$\cdot$10$^6$  &3$\cdot$10$^6$&  & \\
M [M$_\odot$] & 0.20  &0.16&  & \\
M$_B$/M$_A$ & \multicolumn{2}{c|}{0.80} & \multicolumn{2}{c|}{} \\
E$_{H\alpha}$ [nm] & -5.6  &-3.7& -10.9 & -17.8 \\
E$_{LiI}$ [pm] & 110  &61& 24 & 12 \\
lg N$_{Li\,I}$ & -8.2? & -8.8? & -9.8? & ? \\
\end{tabular}
\end{center}
\end{table*}

The highly uncertain distance estimates for the T Tauri stars pose
another problem.  For Chamaeleon I, e.g., the distance estimates given in the
literature range from 115 pc (Th\'e et al.\ 1986) to 220 pc
(Gauvin \& Strom 1992). Similar discrepancies in the distance
estimates exist for Chamaeleon II and $\rho$ Ophiuchi. In the following we
adopt for Chamaeleon I and $\rho$ Oph a distance of 150 pc. For Chamaeleon II
a distance of 200 pc will be assumed (Hughes \& Hartigan 1992) as
a distance as small as 150 pc would not be conform with the high Lithium
abundance observed in the components of Sz 49 and Sz 59 (see below).
As both components of a binary star are at the same
distance, the error in the absolute distance estimate does not
affect the ratio between the estimated luminosities of the two
binary components.

As Figure \ref{sz24_cmd} suggests, extinction values might be variable. 
Therefore estimates of the extinction derived at different epochs cannot be
compared directly. Interestingly, the values derived by us are lower
than the values quoted in the literature for all stars.
{\it The presence of the secondary causes the
unresolved binary to appear redder than the primary actually is and therefore
leads to an overestimate of the line of sight extinction}.
On the other hand, neglecting excess emission due to veiling leads to an 
underestimate of the foreground extinction (Hartigan et al.\ 1991). Therefore,
the next step will be to determine the typical veiling for the individual
components of all the binaries in our sample.

\section{Physical properties: Age, mass, equivalent width, visual extinction,
Lithium abundance}

\begin{figure*}[ht]
\centerline{\vbox{\psfig{figure=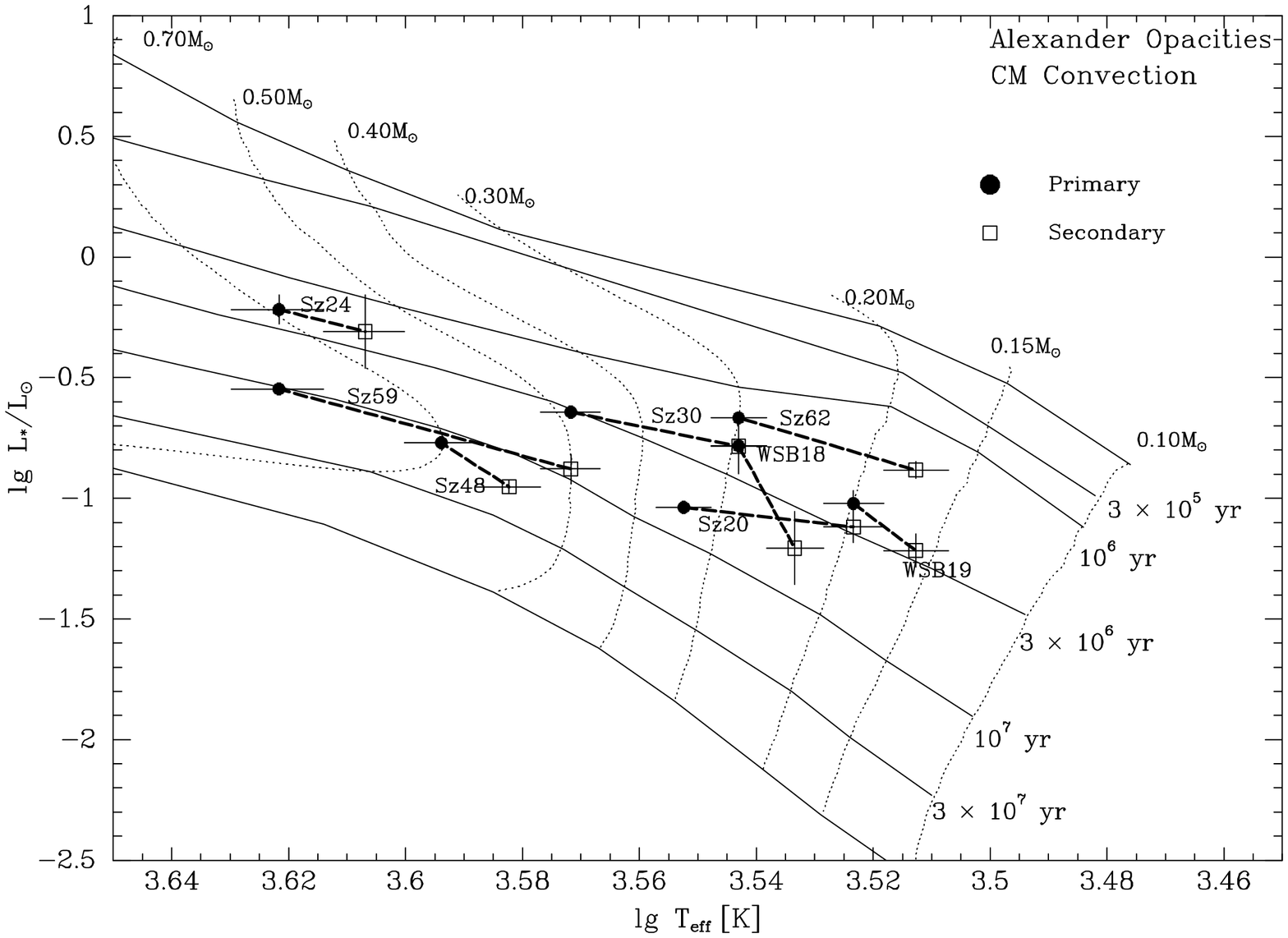,width=16.0cm,angle=90}
            \hbox{\psfig{figure=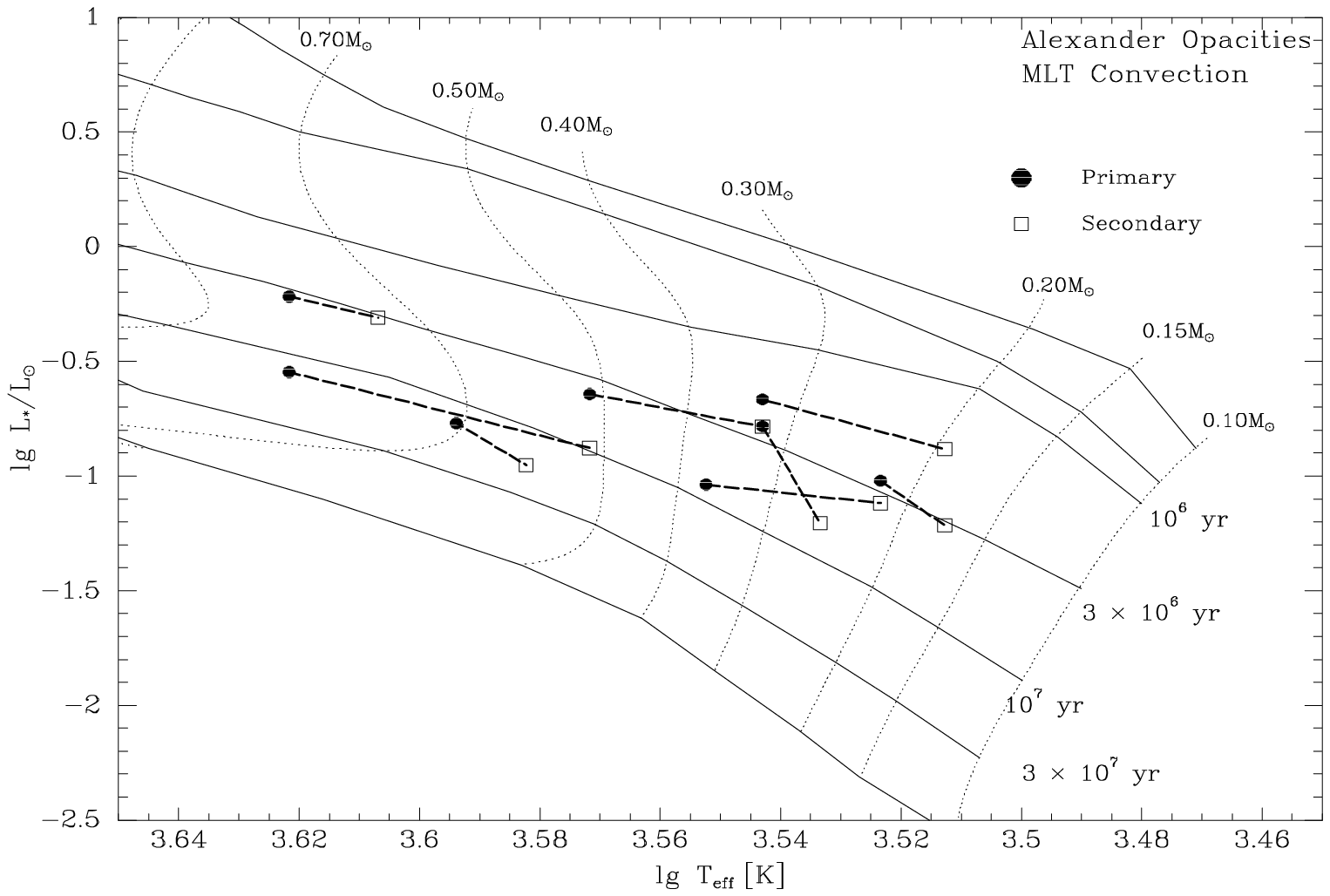,width=8.0cm,angle=0}
                  \psfig{figure=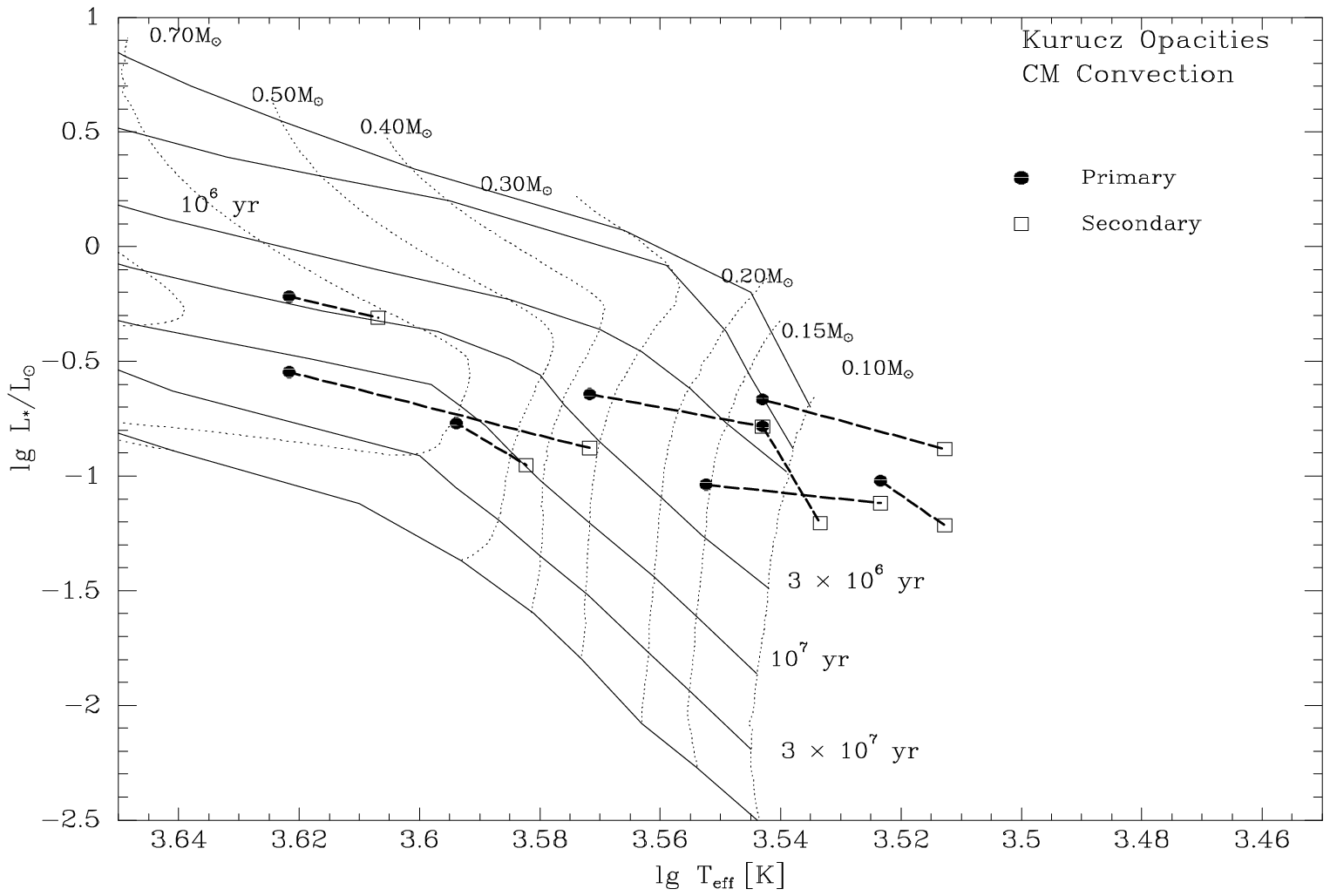,width=8.0cm,angle=0}}}}
\caption{\label{hrd_alex_cm}
Binary components placed on an H-R diagram.  For comparison
theoretical pre-main-sequence evolutionary tracks from D'Antona
\& Mazzitelli (1994) are overplotted (top).
Tracks based on mixing length theory for a description of the
convection can only be distinct from those based on the
Canuto \& Mazzitelli description by a determination
of the dynamical masses of pre-main-sequence binaries (bottom left).
Tracks based on the opacities from Kurucz (1991) do not
provide an adequate description for late-type stars because
of the lack of molecular opacities (bottom right).
}
\end{figure*}

\begin{figure*}[ht]
\centerline{\psfig{figure=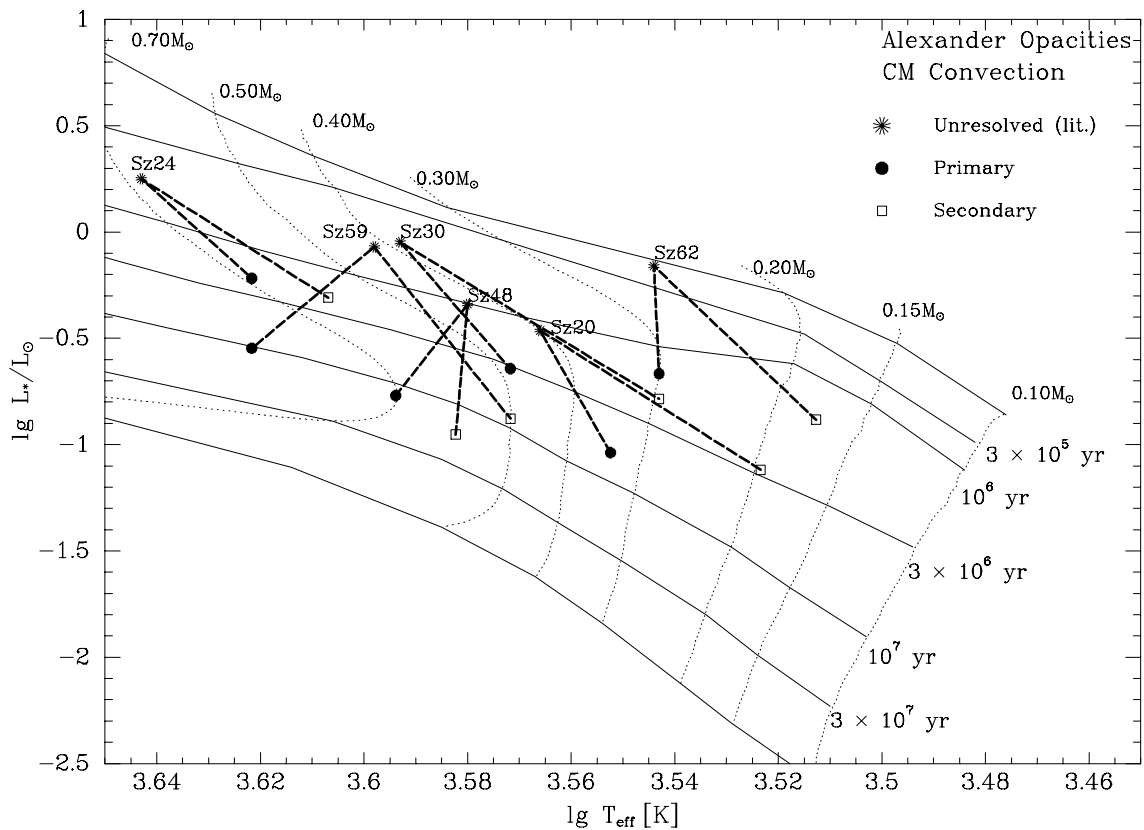,width=16.0cm,angle=90}}
\caption{\label{hrd_unres}
Unresolved binaries lead to an underestimate of the age
of a T Tauri star population. We compare recent literature values
for the unresolved binaries (Feigelson et al.\ 1993, Hughes \& Hartigan 1992)
with our results for the individual binary components.
}
\end{figure*}

Table \ref{tthahrd} summarizes the physical properties of the binary components.
In Figure \ref{hrd_alex_cm} we have placed those binary components in
H-R diagrams for which we could derive
both spatially resolved photometry and spectroscopy.
For comparison we show theoretical pre-main-sequence evolutionary
tracks from D'Antona \& Mazzitelli (1994). These tracks are based
on low-temperature opacities from Alexander et al.\ (1989) or Kurucz (1991) 
and on the convection model by Canuto \& Mazzitelli (1990) or mixing-length
theory. The errors in the
placement of the individual binary components are indicated. The
theoretical tracks allow us to determine the age and mass of the stars.
All of our binaries have mass ratios between 0.5 and 1 (cf.\ Table 
\ref{tthahrd}) whereas only
60\% of the binaries studied by Hartigan et al.\ (1994) have mass ratios
in that range. Hence our sample appears to be biased somewhat towards
equal mass pairs. 
However, even among the 15 binaries with mass ratios
between 0.5 and 1 in the Hartigan et al.\ (1994) sample only eleven (75\%)
are coeval.

Sets of theoretical PMS evolutionary tracks based on different input physics
are shown in Figure \ref{hrd_alex_cm}, bottom. 
A different convection model (i.e.\ mixing length 
theory) affects mainly the effective temperature of the evolutionary 
tracks (and hence the mass estimates) for stars around 0.5 M$_\odot$ to 0.7 
M$_\odot$. In the near future dynamical mass determinations for PMS binaries 
will allow us to distinguish observationally between these two
sets of tracks. Tracks based on opacities from Kurucz (1991) lack molecular 
opacities and hence provide no adequate description for young late-type
stars.

Hartigan et al.\ (1994) compared their observations also to
tracks by Swenson et al.\ (unpublished). Applying the Swenson
tracks to our sample of PMS binaries yield masses above 0.4M$_\odot$
and ages between 3$\cdot$10$^6$ yr and 3$\cdot$10$^7$ yr for the
binary components.
These ages appear to be rather high given the fact that the stars
in our sample are still associated with the molecular clouds (e.g.\ cf.\
Brandner et al.\ 1996; Fig.\ 5 \& 6). 
On the other hand, observations of the spectroscopic binary
NTT 155913-2233 suggest that the tracks by Swenson et al.\
might provide a better description for PMS stars in a mass range
between 0.6M$_\odot$ and 1.1M$_\odot$ than the tracks by
D'Antona \& Mazzitelli (Prato \& Simon, priv.\ comm.).

Figure \ref{hrd_unres} nicely illustrates how age and mass estimates can
be wrong due to unresolved binaries. For
the objects studied in Chamaeleon I and II we have plotted both the
most recent literature values for effective temperature and luminosity 
for the unresolved binaries (Feigelson et al.\ 
1993; Hughes \& Hartigan 1992) and the effective temperatures and
luminosities as derived by us
for the individual binary components. The literature values given by
Feigelson et al.\ (1993) have been scaled to a distance of 150 pc.
Note that the somewhat smaller values for the visual extinction towards the 
binaries derived by us lead also to somewhat lower estimates for the
system luminosity in comparison to Feigelson et al.\ (1993) and 
Hughes \& Hartigan (1992).

Unresolved PMS binaries could lead to a gross underestimate of the age of the
stars involved (cf.\ Ghez 1994) and also induce errors in the individual mass 
estimates and hence in the derivation of the initial mass function. 

\begin{figure*}[ht]
\centerline{\vbox{\hbox{\psfig{figure=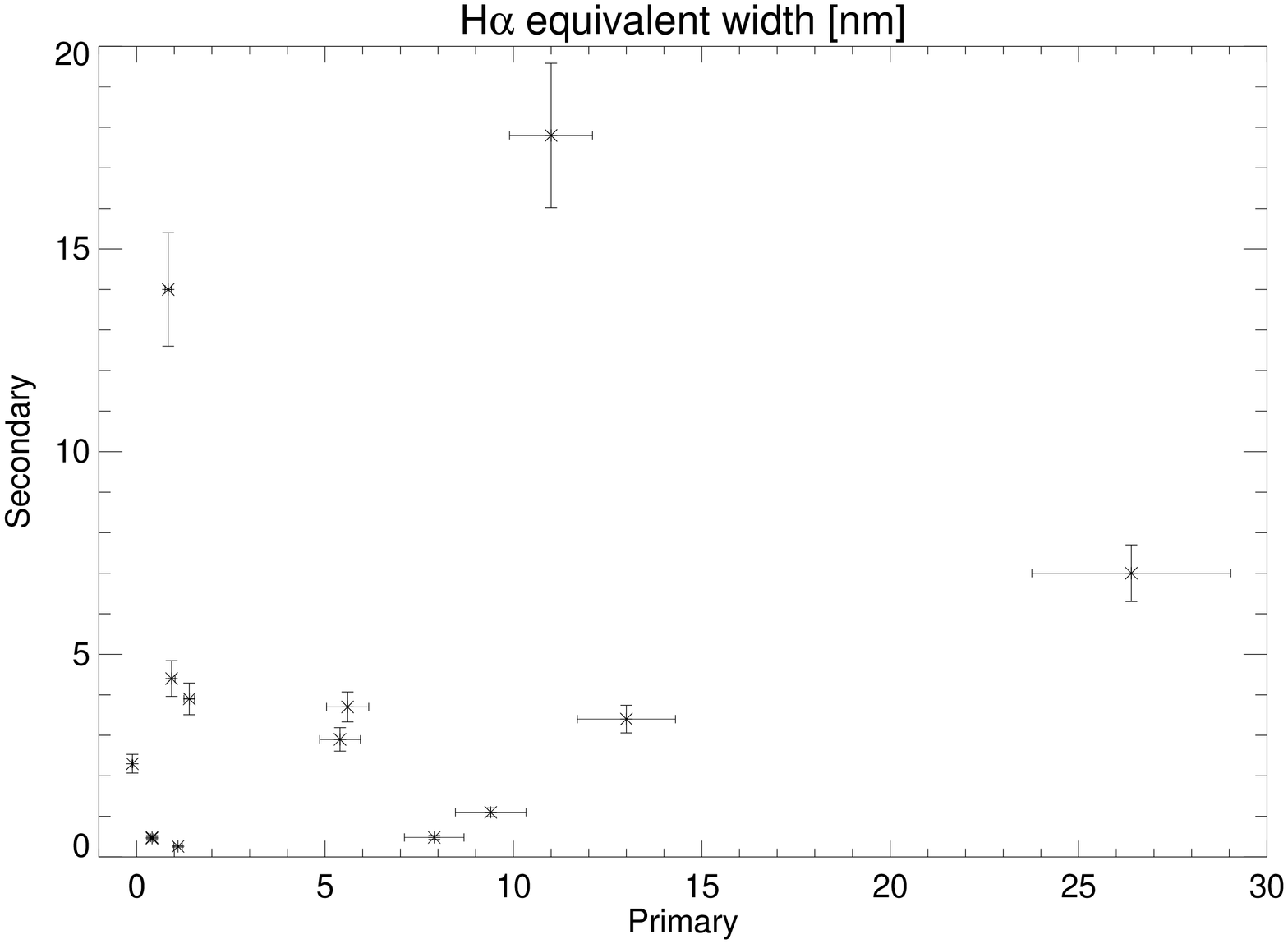,width=8.0cm,angle=90}
                  \psfig{figure=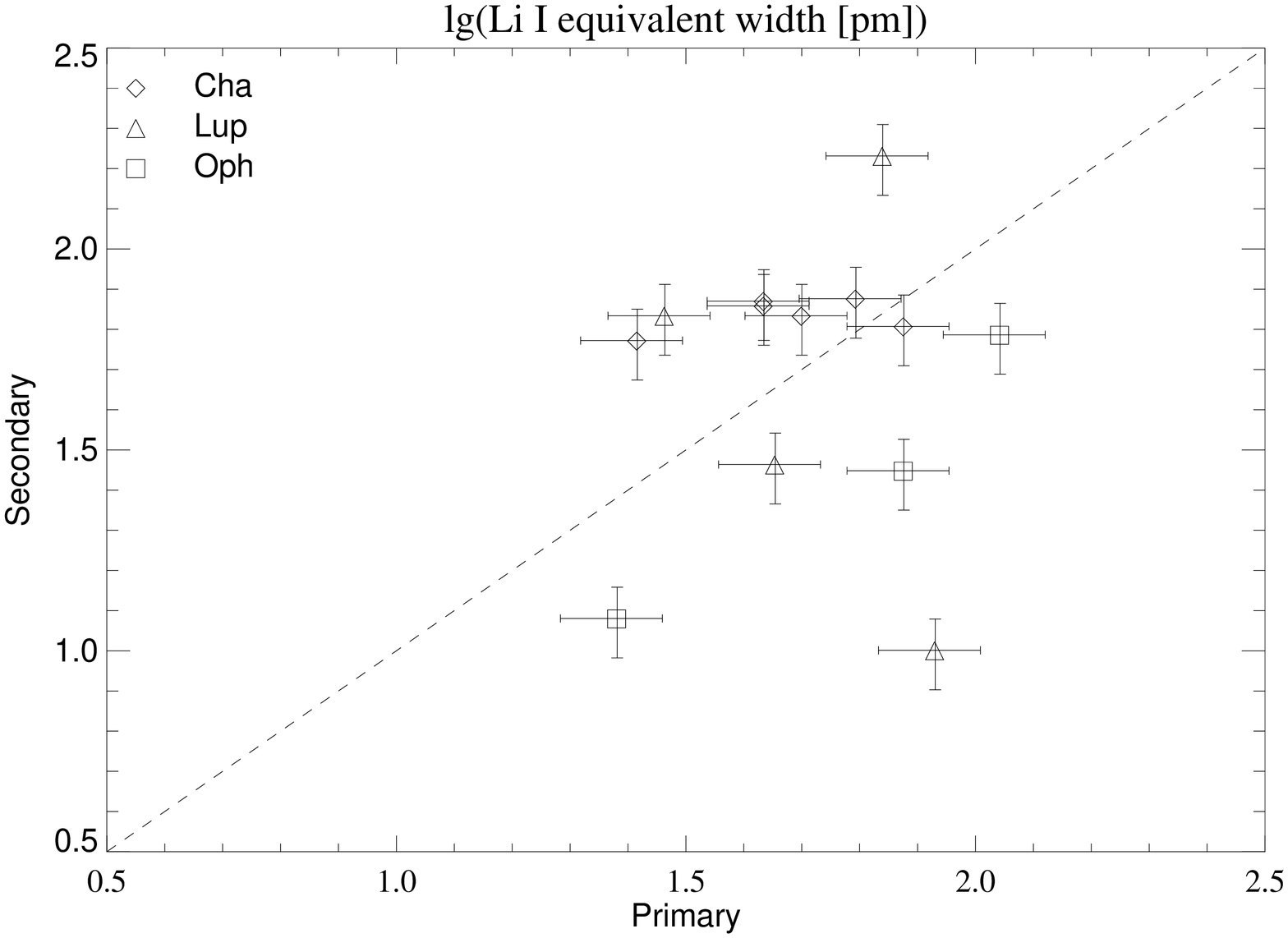,width=8.0cm,angle=90}}
                  \hbox{\psfig{figure=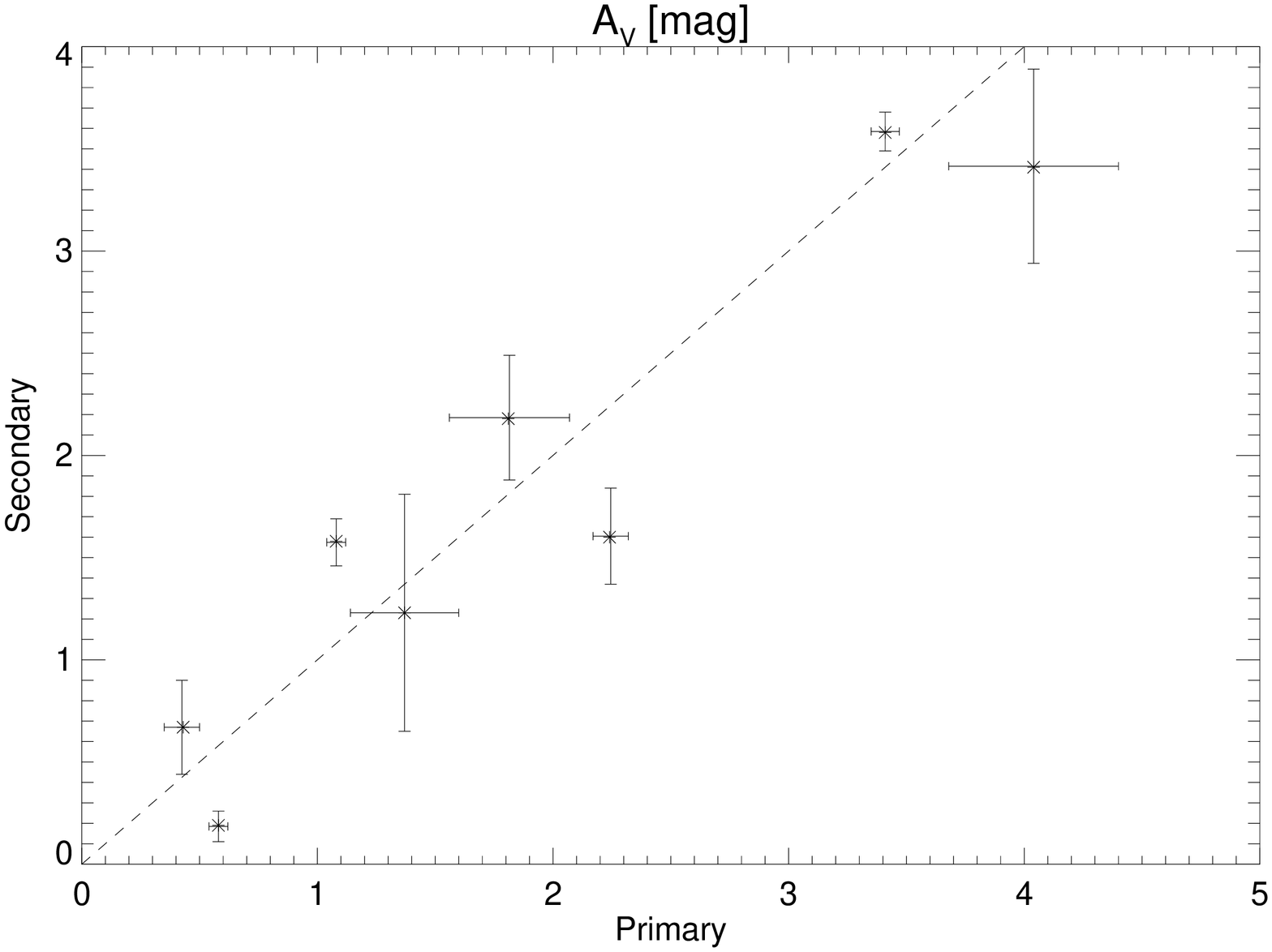,width=8.0cm,angle=90}
                  \psfig{figure=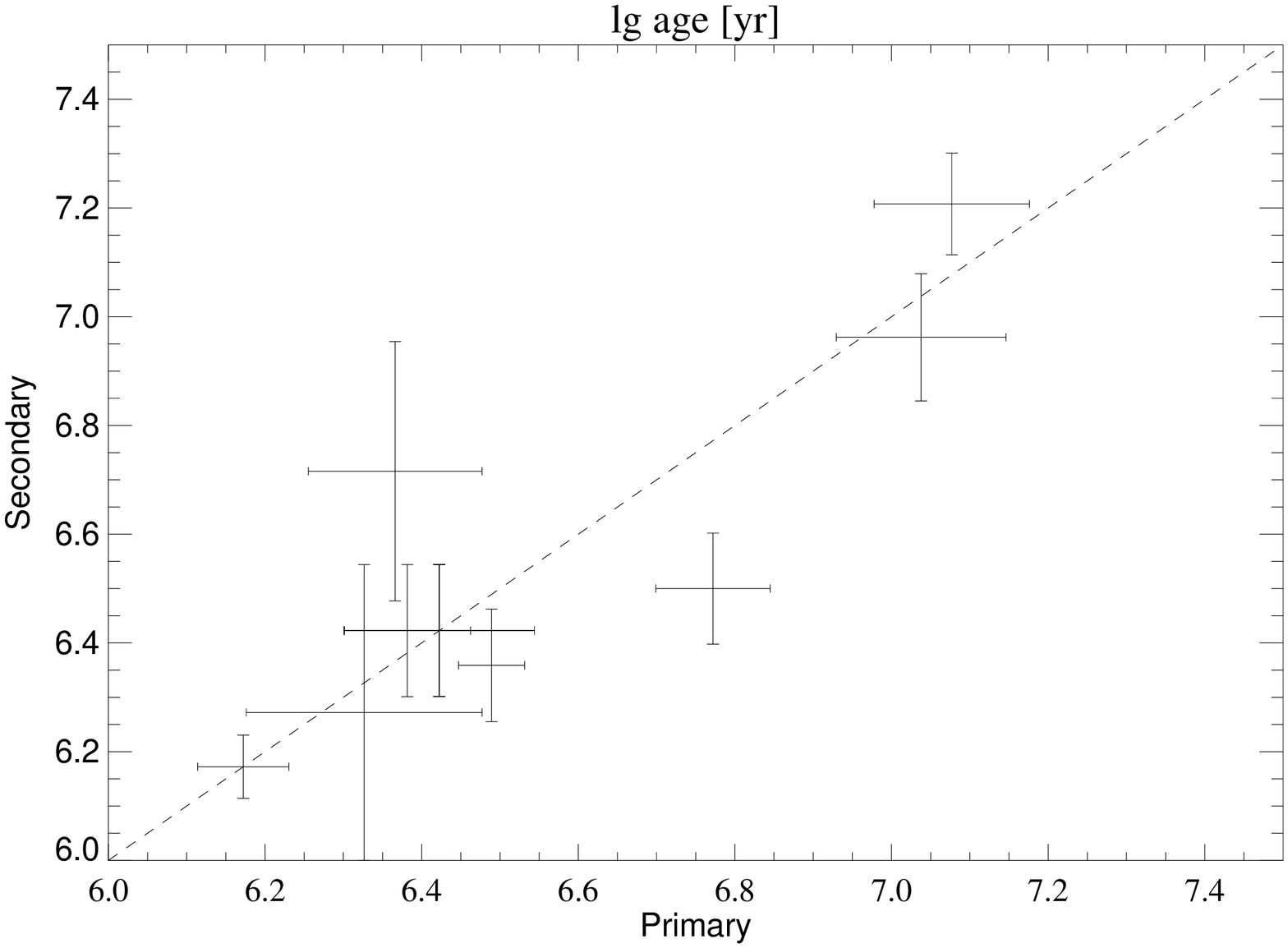,width=8.0cm,angle=90}}}}
\caption{\label{phys_par}
Correlation of physical parameters of primaries and secondaries:
a) H$\alpha$ equivalent widths, which are a measure of the accretion
rate appear not to be correlated. b) Lithium I equivalent widths
are also not correlated.
c) A$_V$: The visual extinction towards the primary and towards the secondary 
of each individual binary show a good correlation.  
d) Age: All binaries appear to be coeval within the statistical errors.}
\end{figure*}

In Figure \ref{phys_par} different physical parameters of the primary and
secondary are plotted against each other. Fig.\ \ref{phys_par}a (top left)
shows the equivalent width of the H$\alpha$ emission line of the
primary and the secondary. No clear correlation exists. Hence,
{\it chromospheric activity and/or accretion rates of primary and
secondary appear not to be related}. There are a number of
binaries in which only the secondary exhibits strong H$\alpha$ emission
but not the primary. Such stars have only been detected
in H$\alpha$ objective prism surveys because they are binaries
and have a secondary bright in H$\alpha$. This selection effect may
explain at least partially why the number of binaries among pre-main-sequence 
stars appears to be higher than among main-sequence stars
(cf.\ Ghez et al.\ 1993; Leinert et al.\ 1993; Reipurth \& Zinnecker 1993).

The Lithium I (670.7nm) equivalent width of primary and secondary 
(Fig.\ \ref{phys_par}b, top right) are not correlated as a test based on rank 
statistics indicates. Differences in the individual veiling and the non-linear 
dependence of the Lithium abundance on age and effective temperature might
be responsible for that.
See below for a more detailed analysis 
considering curve of growth calculations and Lithium depletion as a function 
of age and effective temperature. ESO H$\alpha$ 281 A is the only star in our 
sample of 28 binary components which does not show any sign of Lithium
absorption. The small value of its A index indicates that ESO H$\alpha$ 281 A
is a background giant.

The visual extinction towards the primary and secondary (Fig.\ \ref{phys_par}c, bottom left)
are in good agreement with each other. This indicates that both components
of each binary are embedded equally deep in the dark clouds and thus
gives further evidence that the objects studied are indeed physical
binaries. Furthermore, the agreement of the extinction values yields
that the circumstellar disks around the individual components in each
binary system in our sample are aligned.

\begin{figure}[ht]
\centerline{\psfig{figure=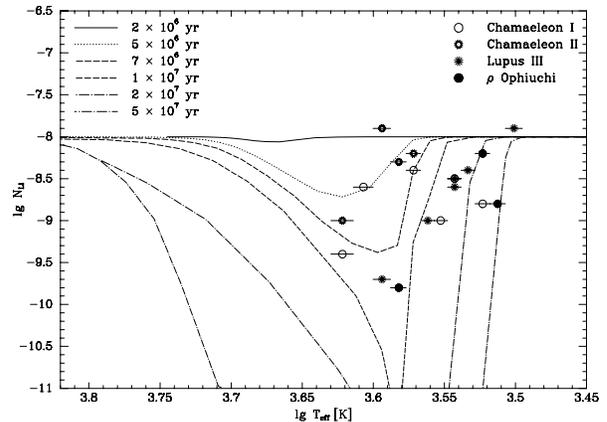,width=8.0cm,angle=90}}
\caption{\label{li_iso}
 Lithium abundance as a function of effective temperature. We have
also plotted theoretical isochrones from D'Antona \& Mazzitelli (1994)
based on opacities from Alexander et al.\ (1989) and the
Canuto \& Mazzitelli (1991) description of convection.
}
\end{figure}

The age of the primary and secondary (Fig.\ \ref{phys_par}d, bottom right) 
derived from the theoretical
pre-main-sequence evolutionary tracks is generally in good agreement
for ages in a range of a few 10$^6$ yr to 10$^7$ yr.
Only for one star (Sz 20) the secondary might be somewhat younger than
the primary. However, the deviation amounts to only 1.5 $\sigma$ and thus
could very well be a purely statistical fluctuation.

Based on the curve of growth calculations by Mart\'{\i}n et al.\
(1994) we were able to derive Lithium abundances for
our sample of pre-main-sequence binary components. Figure \ref{li_iso}
shows the Lithium abundances as a function of effective temperature.
Overplotted are theoretical isochrones from
D'Antona \& Mazzitelli (1994). As the curve of growth calculations are
based on Kurucz model atmospheres they are only valid down to effective
temperatures of about 3700 K. Below that temperature the abundances
derived from the model calculations lie probably too low
(Mart\'{\i}n et al.\ 1994). Veiling
of photospheric lines could additionally lead to an underestimate of the
actual Lithium abundance. Hence, the values plotted in Figure \ref{li_iso} merely
represent lower limits in most cases.

For the components of Sz 48 and Sz 59 the Lithium isochrones suggest
an age of less than 10$^7$ yr. In order to match this with their positions
in the H-R diagram the distance to these stars (and thus towards the
Chamaeleon II cloud) has to be at least 200 pc. 

\section{Summary}

We have surveyed 14 binaries with separations between 0\farcs6 and
1\farcs7. In 27 of the 28 individual stars we did find Lithium absorption,
which (together with their H$\alpha$ emission and association to
dark clouds) classifies them as T Tauri stars. 
One of the presumed binaries turned out to be a likely chance projection
with the ``primary'' showing neither H$\alpha$ emission nor Lithium absorption.
This object (ESO H$\alpha$ 281 A) is very likely a background giant.
Similarly, Aspin et al.\ (1994) found the `companion' to the
PMS star ESO H$\alpha$ 279 A to be a background giant.

A comparison of the equivalent width of the H$\alpha$ emission line
of the primaries and secondaries showed that they are not
correlated with each other. Some of the originally unresolved
T Tauri stars were only picked up in H$\alpha$ surveys because they
are binaries and have a {\it secondary} with strong H$\alpha$ emission
whereas the primary shows only weak H$\alpha$ emission.
Therefore, samples of H$\alpha$ selected T Tauri
stars might be biased towards binaries. 

Eight out of the 14 pairs could be placed on an H--R diagram. A comparison
with theoretical pre-main-sequence evolutionary tracks yields that for
{\it all} pairs the individual components appear to be coeval within
the observational errors. 
This finding is similar to Hartigan et al.\ (1994) who
found that 2/3 of the wider pairs with separations from 400 AU to
6000 AU are coeval.
However, unlike Hartigan et al.\ for the wider pairs, we find no non-coeval
pairs among our sample. Thus, young binaries with separations less than 400 AU 
might indeed represent a different regime as compared to binaries with 
separations between 400 AU and 6000 AU. Otherwise, it might very well be
that the wide non-coeval binaries in the sample from Hartigan et al.\
are just chance projections and not physically related to each other.

\acknowledgements
We are grateful to Bo Reipurth for his help in preparing one of the
proposals. We would like to thank France Allard for communicating
the synthetic colours derived from the model spectra of M dwarfs to us.
We are grateful to Pat Hartigan for his helpful  referee's comments.
WB was supported by a student fellowship of the European Southern
Observatory and by the Deutsche Forschungsgemeinschaft (DFG) under grant
Yo 5/16-1. 
This research has made use of the Simbad database,
operated at CDS, Strasbourg, France, and NASA's Astrophysics Data System
Abstract Server.


\begin{thebibliography}{}

\bibitem{} Alexander D., Augason G., Johnson H.\ 1989 ApJ 345, 1014

\bibitem{} Allard F., Hauschildt P.H.\ 1995 ApJ 445, 433

\bibitem{} Allen L.E., Strom K.M.\ 1995 AJ 109, 1379

\bibitem{} Appenzeller I.\ 1977 A\&A 61, 21

\bibitem{} Appenzeller I.\ 1979 A\&A 71, 305

\bibitem{} Appenzeller I., Krautter J., Jankovics I.\ 1983 A\&AS 53, 291

\bibitem{} Artymowicz P., Lubow S.H.\ 1994 ApJ 421, 651

\bibitem{} Aspin C., Reipurth B., Lehmann T.\ 1994 A\&A 288, 165

\bibitem{} Bertout C., Basri G., Bouvier J.\ 1988 ApJ 330, 350

\bibitem{} Bodenheimer P., Rozyczka M., Yorke H.W., Tohline J.E.\ 1988
in {\it Formation and Evolution of Low Mass Stars} p.\ 139, eds.\ Dupree A.K.\
\& Lago M.T.V.T.

\bibitem{} Boss A.P.\ 1988 Comments in Astrophys.\ 12, 169

\bibitem{} Bouvier J., Bertout C., Bouchet P.\ 1988 A\&AS 75, 1

\bibitem{} Brandner W.\ 1992 Diploma thesis, Universit\"at W\"urzburg 

\bibitem{} Brandner W., Alcal\'a J.M., Kunkel M., Moneti A., Zinnecker H.\
1996 A\&A 307, 121

\bibitem{} Canuto V., Mazzitelli I.\ 1990 ApJ 370, 295

\bibitem{} D'Antona F., Mazzitelli I.\ 1994 ApJS 90, 467

\bibitem{} Feigelson E.D., Casanova S., Montmerle T., Guibert J.\ 1993 ApJ 416, 623

\bibitem{} Gauvin L.S., Strom K.M.\ 1992 ApJ 385, 217

\bibitem{} Ghez A.\ 1994 BAAS 185, 99.01

\bibitem{} Ghez A., Neugebauer G., Matthews K.\ 1993 AJ 106, 2005

\bibitem{} Grinin V.P.\ 1992 A\&AT 3, 17

\bibitem{} Hartigan P.\ 1993 AJ 105, 1511

\bibitem{} Hartigan P., Kenyon S.J., Hartmann L.\ et al.\ 1991 ApJ 382, 617

\bibitem{} Hartigan P., Strom K.M., Strom S.E.\ 1994 ApJ 427, 961

\bibitem{} Hughes J., Hartigan P.\ 1992 AJ 104, 680

\bibitem{} Hughes J., Hartigan P., Krautter J., Kelemen J.\ 1994 AJ 108, 1071

\bibitem{} Jeffrys W.H., McArthur B., McCartney J.E.\ 1991 BAAS 23, 997

\bibitem{} Kenyon S.J., Hartman L.W.\ 1990 AJ 99, 869

\bibitem{} Kirkpatrick J.D., Kelly D.M., Rieke G.H.\ et al.\ 1991 ApJS 77, 417

\bibitem{} Kurucz R.\ 1991 in Crivellari \& Hubeny (eds.), NATO ASI Series, p.\ 441

\bibitem{} Larson R.B.\ 1978 MNRAS 184, 69

\bibitem{} Leinert C., Zinnecker H., Weitzel N., et al.\ 1993 A\&A 278, 129

\bibitem{} Mart\'{\i}n E., Rebolo R., Pavlenko A.M.Y.\ 1994 A\&A 282, 503

\bibitem{} Mundt R., Bastian U.\ 1980 A\&AS 39, 245

\bibitem{} Papaloizou J.C.B., Pringle J.E.\ 1977 MNRAS 181, 441

\bibitem{} Pringle J.E.\ 1989 MNRAS 239, 361

\bibitem{} Reipurth B., Zinnecker H.\ 1993 A\&A 278, 81

\bibitem{} Rydgren A.\ 1980 AJ 85, 444

\bibitem{} Schwartz R.D.\ 1977 ApJS 35, 161

\bibitem{} Th\'e P., Wesselius P.R., Tjin-A-Djie H.R.E., Steenman H.\ 1986 A\&A 155, 347

\bibitem{} Turnshek D.E., Turnshek D.A., Craine E.R., Boeshaar P.C.\ 1985 Technical Report,
Western Research Company, Arizona

\bibitem{} Walter F.M., Vrba F.J., Mathieu R.D., Brown A., Myers P.C\ 1994 AJ 107, 692

\bibitem{} Wilking B.A., Schwartz R.D., Blackwell J.H.\ 1987 AJ 94, 106

\end{thebibliography}
\end{document}